# Adaptive Fault Masking With Incoherence Scoring

Barış Baykant Alagöz

**Abstract**: *An adaptive voting algorithm for digital media was introduced in this study. Availability was improved by incoherence scoring in voting mechanism of Multi-Modular Redundancy. Regulation parameters give the algorithm flexibility of adjusting priorities in decision process. Proposed adaptive voting algorithm was shown to be more aware of fault status of redundant modules.*

## Introduction:

There has been several works on Fault Detection and Isolation (FDI).[1],[2] Multi Modular Redundancy (Multi-MR) was a widely used technique for fault-masking. In a Multi-MR system [6]-[8], we have more than one implementations of the same logic function and their outputs are voted using a voter circuit.(see Figure 1) The winner output is selected as system outputs. Triple Modular Redundancy (TMR) is a widely used redundancy technique for fault masking.[9]-[11] Word-Voter technique [5] and bit-by-bit voting schemes were used in TMR. Adaptive majority voting algorithm utilizing reliability history of modules was developed. [12]

In this study, we proposed improved fault-masking capability for digital media by developing an algorithm based on adaptive voting technique. We suggested an Adaptive Multi Modular Redundancy mechanism, which was benefiting from regulatable cooperation between conventional bit-by-bit majority voting of redundant modules and their incoherence history. The incoherence history of a redundant module comprises weighted hamming distances between adaptive voting outputs and output of module itself. Thus, adaptive voting algorithm utilizes not only coherence at the outputs of modules, but also, benefits from incoherence history of the digital modules, which was updated after every decision process of adaptive voter. Incoherence history indeed constitutes a memory, which holds discordance between voter decisions and module outputs. This property gains the adaptive voting algorithm skill of learning from the past experiences with redundant modules. The other important point to be considered is that adaptive voting algorithms do not need any external error detection [4] and measurement mechanism.

In previous version of adaptive voting algorithm, the most reliable modules can be selected if attending a majority consensus. [12] So, reliable result must first be in majority group and then it has to be the most reliable member in majority group in order to be selected. Hence, being in majority consensus has priority in decision process of selecting the most reliable modules. In our study, we were proposed a flexible adaptive voter behavior controlled by two regulation parameters ($\alpha,\beta$). These regulation parameters determine weight of majority consensus and weight of reliability background of modules (incoherence history) in decision process. These regulation parameters (α,β) give the adaptive voting algorithm capability of tuning priorities in decision process, instantaneously.

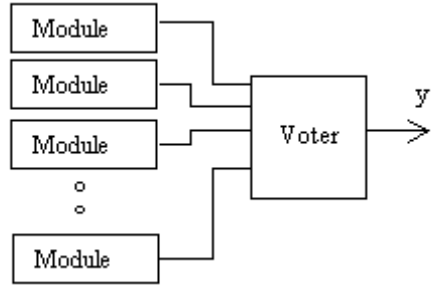

**Figure 1**. Multi Modular Redundancy

We have observed that proposed adaptive voting algorithm could be more aware of fault status of redundant modules than majority voter. Therefore, it exhibits better performance of fault masking when the fault free modules exist in multi-modular structure.

## Adaptive Voting Algorithm Based On Incoherence:

a) Introduction Of Algorithm

Lets $k$ denote number of redundant modules outputs, which are $y_1, y_2, ..., y_k$ and let $y$ represent adaptive voter output. For the digital system, redundant modules were assumed to be Boolean functions. Normalized hamming distance, which is defined as division of the number of different bits in two logic words to number of bits in a logic word, was preferred for the measure of incoherence.

$$Inc(y_i, y_j) = \frac{HamDis(y_i, y_j)}{m} \quad (1)$$

Here, $Inc(y_i, y_j)$ is incoherence between output $y_i$ and output $y_j$ and $m$ number of bits in $y_i$. Incoherence history for output $y_i$ at the moment of $n$ was represented by $Rs(y_i, n)$ state variable.

$$Rs(y_i, n+1) = \alpha.Inc(y_i(n), y(n)) + (1-\alpha).Rs(y_i, n) \quad (2)$$

Here, $Rs(y_i, n+1)$ is incoherence history of module output $y_i$ for the next processing and $n$ is time index. $Rs(y_i, n)$ is current value of incoherence history of module output $y_i$, correspondingly. Normalized hamming distance between module output $y_i$ and adaptive voter decision $y$ was used for updating incoherence history $Rs$. The parameter of α is memory regulation parameter and takes value in range of 0.0 to 1.0. Higher α enforces short-term memory behavior and lower α results a long-term memory. As seen equation 1, lower bound of incoherence measure $Inc(y_i, y)$ will be equal to 0.0 and it is seen when $y_i$ module output was exactly equal to adaptive voter output $y$. Upper bound of the incoherence measure $Inc(y_i, y)$ will be 1.0 and it is seen when all bits of $y_i$ and $y$ were different from each other. Incoherence history $Rs$ represented by (2) has an upper limit of 1.0 for the sake of factors α and (1-α). Equation (2) yields incoherence history to be used for the next running of adaptive voting algorithm. Incoherence score is given as following,

$$Is(y_i, n) = \beta.Inc(y_i, y_c) + (1-\beta).Rs(y_i, n) \qquad (3)$$

Here, $Is(y_i, n)$ represents incoherence score and β is impulsiveness regulation parameter and takes value in range of 0.0 to 1.0. Higher value of β increases weight of majority voting. Lower value of β increases weight of incoherence history in decision process. $y_c$ represents majority voting result for current redundant outputs $y_1, y_2,..., y_k$. Adaptive voting algorithm makes its decision by selecting redundant module output, which has minimum incoherence score. Then, it sets it to output $y$ as the most reliable output of the adaptive voter.

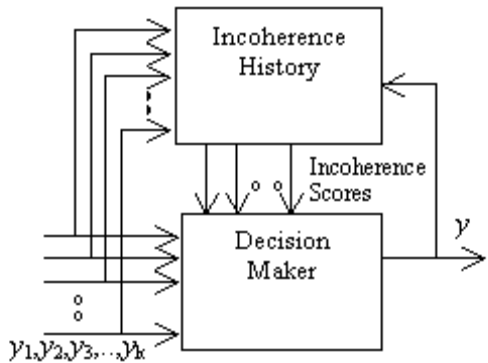

**Figure 2**. Block diagram of adaptive voter with incoherence score

In the Figure 2, block diagram of the adaptive voting was illustrated to describe process. Decision maker block, decides the most reliable output by considering minimum incoherence score. Adaptive Voting algorithm performs following algorithm steps:

**Step 1**: Calculate $y_c$ majority voting output for modules outputs $y_1, y_2, ..., y_k$

**Step 2**: Calculate incoherence scores $Is(y_i, n)$ for all outputs $y_1, y_2, ..., y_k$.

**Step 3**: Select minimum incoherence scores $Is(y_i, n)$ and set corresponding $y_i$ to adaptive voter output $y(n)$

$$y(n) = \{y_i \quad \min(Is(y_i(n), n)) \wedge i \in [1, k]\} \quad (4)$$

**Step 4**: Update incoherence history $Rs(y_i)$ for all redundant module outputs $y_1, y_2, ..., y_k$ according to (2) and go to step 1.

This algorithm is independent of functionality of redundant modules and it makes the algorithm versatile in use. We particularly researched fault masking in digital media.

b) Effects of Regulation Parameter (α,β) and Dynamic Setting

Regulation parameters (α,β) specify priorities in decision process of adaptive voting algorithm. Change in attitude of adaptive voter depending of (α,β) was illustrated in diagram seen in Figure 3. β parameter is impulsiveness parameter. Increasing β enforces giving priority in decision to majority voter and results an impulsive nature in character of adaptive voter. Lower β makes adaptive voting wiser by increasing weight of incoherence history in decision process.

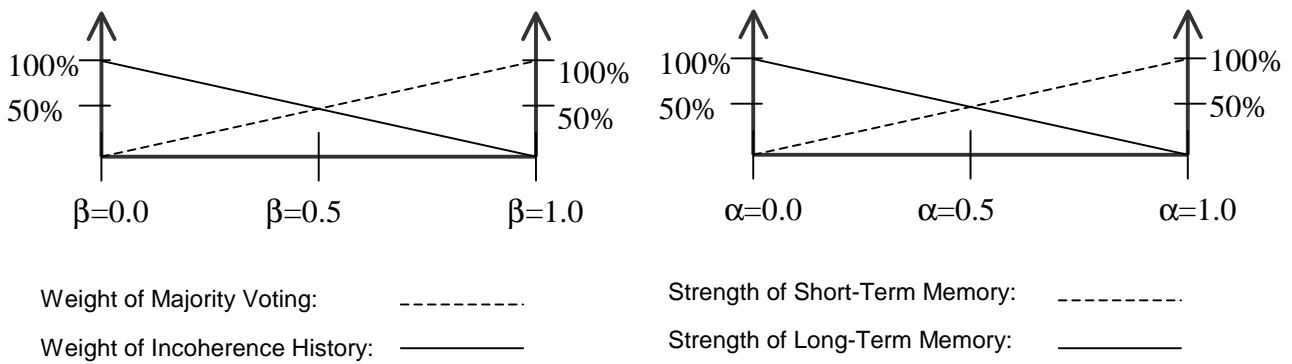

**Figure 3**. Effect of the regulation parameters on decision process

α parameter is memory regulation parameter. Increasing α strengthens short-term memory and increase importance of recent faults in decision process. Lower α leads adaptive voter poses long-term memory that rise consideration of much older experience with redundant modules.

Setting regulation parameters ($\alpha,\beta$) to a low value makes adaptive voter rely on incoherence history and results more aware of the fault-free modules. This property helps better masking of faulty modules as long as there still exists a fault-free module among redundant modules. When all redundant modules were faulty, relying on majority voting could give better results. These two statements help us develop dynamic adaptive voting algorithm that can automatically change regulation parameters ($\alpha,\beta$) to improve performance of fault masking. Rule for two state dynamic regulation parameter setting can be derived as following:

**1-** If there exist at least one fault-free module, rely on incoherence history. So, set a low value to $\beta$.
**2-** If all redundant modules diagnosed as faulty, rely on majority voting. So, set a high value to $\beta$.

In the Figure 4, a software implementation for dynamic setting of the regulation parameter was illustrated. In this implementation, $Rs(y_i,n)$ parameter of each module was used for diagnostic proposes to determine the fault status of redundant modules. Being equal or lower than a threshold indicates corresponding redundant module to be fault-free. Otherwise, redundant module diagnosed as faulty. Program seen in Figure 4 set $\beta$ regulation parameter to a high value, when it diagnosed all redundant modules faulty.

```
% Vth threshold for diagnostic
Vth=0.001;

% When AllFaultyIndicator variable is true, it indicates
% that all redundant modules diagnosed as faulty.
AllFaultyIndicator = (Rs(y1) >Vth) And (Rs(y2) >Vth)
And (Rs(y3) >Vth) And (Rs(y4) >Vth) And (Rs(y5)
>Vth)

if AllFaultyIndicator == true
   Beta=0.8;
else
   Beta=0.3;
end
```

**Figure 4**. Soft implementation of the dynamic regulation parameter setting

c) Implementation And Application

Soft implementation of the adaptive voter with incoherence scoring for 5 redundant modules is illustrated in the Figure 6. In this program, *m* is the number of bit at the output of redundant modules. As seen in soft implementation, one bit by bit majority voter, one incoherence score $Is(y_i)$ calculation (Equation 3) and one incoherence history update calculation $Rs(y_i)$ are required for every additional redundant modules.

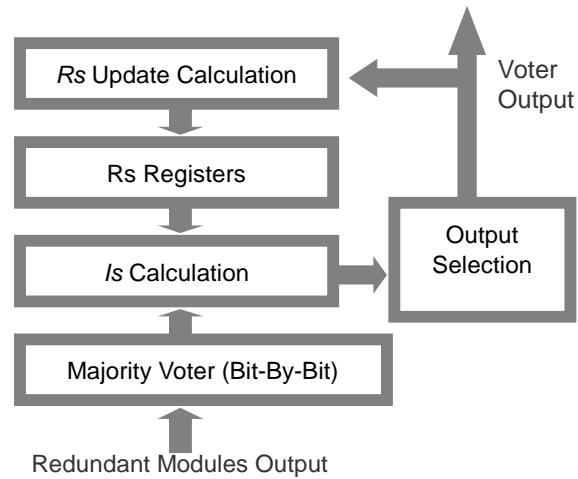

**Figure 5**. Hardware implementation of the dynamic regulation parameter setting

In the figure 5, block diagram for hardware implementation was given to demonstrate complexity of circuitry.

## Performance Comparison With Voting Algorithms:

a) Voting Algorithms Overview

In this section, output selection mechanism of some voting algorithms were discussed to evaluate fault-masking performance.

```
% Algorithm step 1
yc(i)=MajorityVotingBitByBit(y1,y2,y3,y4,y5);

% Algorithm step 2
Is(y1)=Beta*HammingDist(y1,yc)/m
+(1-Beta)*Rs(y1);
Is(y2)=Beta*HammingDist(y2,yc)/m
+(1-Beta)*Rs(y2);
Is(y3)=Beta*HammingDist(y3,yc)/m
+(1-Beta)*Rs(y3);
Is(y4)=Beta*HammingDist(y4,yc)/m
+(1-Beta)*Rs(y4);
Is(y5)=Beta*HammingDist(y5,yc)/m
+(1-Beta)*Rs(y5);

% Algorithm step 3
SelectedModule=SelectMinimum(Is(y1), Is(y2),
 Is(y3), Is(y4), Is(y5));
switch SelectedModule
   case y1,
      AdaptiveVoterOutput=y1;
   case y2,
      AdaptiveVoterOutput=y2;
   case y3,
      AdaptiveVoterOutput=y3;
   case y4,
      AdaptiveVoterOutput=y4;
   case y5,
      AdaptiveVoterOutput=y5;
end

% Algorithm step 4
Rs(y1)=Alfa*HammingDist(y1,
AdaptiveVoterOutput)/m+(1-Alfa)*Rs(y1);
Rs(y2)=Alfa*HammingDist(y2,
AdaptiveVoterOutput)/m+(1-Alfa)*Rs(y2);
Rs(y3)=Alfa*HammingDist(y3,
AdaptiveVoterOutput)/m+(1-Alfa)*Rs(y3);
Rs(y4)=Alfa*HammingDist(y4,
AdaptiveVoterOutput)/m+(1-Alfa)*Rs(y4);
Rs(y5)=Alfa*HammingDist(y5,
AdaptiveVoterOutput)/m+(1-Alfa)*Rs(y5);
```

**Figure 6**. Soft implementation of the adaptive voter with incoherence scoring for 5 redundant modules

### *Majority Voter Using Distance Metric:*

For the $y_1, y_2,..., y_k$ redundant modules, the largest sub-group, which satisfies condition of the $|y_i - y_j| \leq a$ for all member of it, constitutes $V_m$ majority group of redundant modules. $a$ parameter is the consensus threshold. We used hamming distant as distance metric due to digital system. Selected output of the voter from the $V_m$ majority group will be the output $y_i$ that is in minimum distance to other member of $V_m$. To better understand, let see Figure 7 showing mapping of $y_1, y_2,..., y_5$ in hamming space. In this figure, $V_m$ were marked by surrounding circle and selected output was indicated by square including $y_4$. Because, $y_4$ have minimum total distance to other member of the $V_m$.

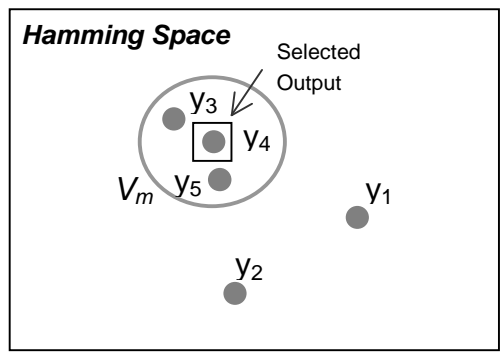

**Figure 7**. Majority voter with word voting decision mechanism

When the all member of the $V_m$ are incorrect, selected output will inevitably be incorrect. It will reduce availability of the voter, in case all redundant modules were faulty. Nevertheless, majority-voting mechanism will reduce error rates in case of being low faulty modules. If there exist one fault-free module, majority voter will not be aware of it to select correct output. This weakness reduces availability of the majority voter compared to adaptive votes when there was a fault-free module. Adaptive voters are more aware of fault-free modules by mean of history record mechanisms.

### *Majority Voter Using Bit-By-Bit Voting Scheme:*

Bit by bit majority voting scheme is the simplest voting scheme to implement in the digital system. Every bit of $y_1, y_2,..., y_k$ are compared each other and voter output $y$ is composed of logic values that has majority in bits of $y_1, y_2,..., y_k$ redundant modules. Figure 8 represents output generation process of bit-by-bit majority voting in hamming space. As clearly seen in Figure 8, bit-by-bit voting can produce output, which is different from $y_1, y_2,..., y_5$. This is why; it has advantages when all of redundant modules outputs were incorrect. Bit-by-bit voting can produce more correct outputs when all modules were faulty. It has exhibited higher availability in the simulations in case all redundant modules were faulty.

Bit-by-bit majority voter isn't aware of one fault-free module like majority voter using distance metric. But, in case all redundant modules output is incorrect, bit-by-bit majority voter can produce more correct results than majority voter using distance metric. Because, majority voter using bit-by-bit voting scheme can produce correct outputs, when error bits at the outputs of redundant modules stay in minority, even if all outputs of modules are incorrect.

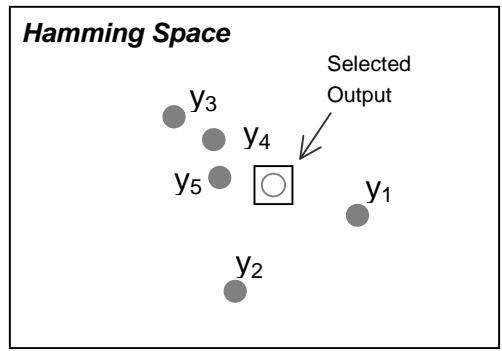

**Figure 8.** Majority voter with bit-by-bit voting decision mechanism

*Adaptive Majority Voter:*

Adaptive majority voter was studied in detail by Latiff and Bennett.[12] In this study, in order to apply adaptive voting algorithm on digital system, hamming distance was used as distance metric. Algorithm selects the most reliable module output from $V_m$ majority group such that redundant module has the largest history record when cardinality of $V_m$ is equal or greater than $(k+1)/2$. While cardinality of $V_m$ is lower than $(k+1)/2$, algorithm produces '*No-Result*'. In the implementation, we assigned majority voter output to adaptive majority voter output in the case of '*No-Result*'. Lets see Figure 9 to demonstrate mechanism of the algorithm. In this figure, value of the history records of modules was coded by gray-level colors. Darker point indicates larger value of history record of the modules. Algorithm selected the darkest point from $V_m$ group in the figure. When the all members of the $V_m$ were incorrect, selected output by adaptive majority voter will be incorrect, correspondingly. But, adaptive majority voter is more aware of fault-free module than majority voters for the sake of history record mechanism. As long as, it is found in $V_m$, adaptive majority voter will possibly select fault-free modules. Therefore, adaptive majority voter can still produce correct results until all redundant modules become faulty in the system.

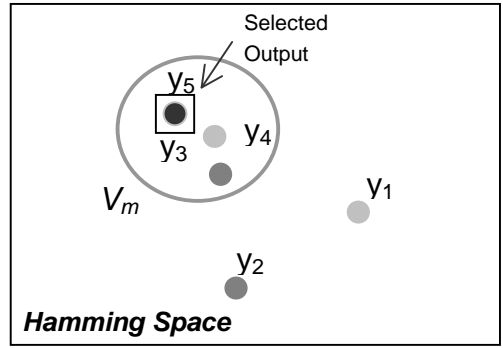

**Figure 9**. Adaptive majority voter decision mechanism

*Adaptive Voter With Incoherence Scoring:*

In the previous section, adaptive voter with incoherence scoring was widely introduced. Basically, it applies both of the incoherence history and majority voter using bit-by-bit scheme in decision process to select the most reliable module. Regulation parameters specify weight of the incoherence history and weight of the majority voter using bit-by-bit scheme in decision process. Depending on regulation parameter, attitude of the voter can vary from wiser, which mostly rely on incoherence history, to impulsive, which mostly rely on majority voter. In the Figure 10, a demonstration of the module selection was illustrated. Both distance to bit-by-bit majority voting result and value of incoherence history can be taken account in selection of modules output. In the figure, redundant module output that has lowest incoherence score ($Is$) was selected.

When all redundant modules outputs were incorrect, adaptive voting with incoherence scoring would select an incorrect output. This will be factor reducing availability of the voter. But, dynamic regulation parameter setting was developed to relieve this weakness. When, voter diagnosed that all modules had become faulty, it increases weight of majority voter in decision process. If it detects a fault-free module available, it will increase weight of incoherence history in decision process. This flexibility of dynamic adaptive voter helps the voting algorithm exhibit better performance under condition that all modules became faulty.

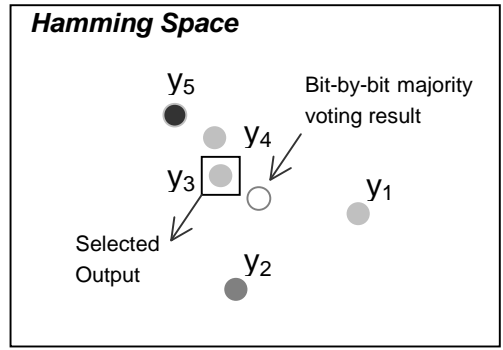

**Figure 10**. Adaptive voter with incoherence scoring decision mechanism

b) Performance Comparisons of Voters In 16-bit Full Adder Circuit Application

In this test scenario, 16-bit full adder circuit was supported by various 5-MR voters and their availability under stuck_one and stuck_zero type permanent fault were compared in Table 1. Availability index of a module is defined as following,

$$A = \frac{n_c}{n} \qquad (5)$$

Here, $n_c$ is number of correct output and $n$ is total number of output. Following test scenario was applied on the redundant modules for realistic simulation,

$$TestSenario_{5MR} = \begin{cases} (N,N,N,N,F), (N,N,N,F,F), \\ (N,N,F,F,F), (N,F,F,F,F), \\ (F,F,F,F,F) \end{cases} \qquad (6)$$

Here, $F$ represents faulty modules and $N$ is fault-free modules. Elements of test scenario represent a test session with modules specified character of faulty ($F$) or fault-free ($N$). In accordance with hardware fails, test of system was performed for every test sessions starting with $(N,N,N,N,F)$ and completed with $(F,F,F,F,F)$ session that it means that all redundant modules became faulty. For every test session in the test scenario, 10.000 random inputs were applied on the full adder modules. Test scenario repeated ten times to make testing more reliable and average of the results, obtained from these test, were reported on the Table 1.

**Table 1.** Availability of the various voters

| System  | Test Scenario of 5-MR | | | | | |
|---------|--------|--------|--------|--------|--------|--------|
|         | NNNNF  | NNNFF  | NNFFF  | NFFFF  | FFFFF  | Totally |
| Module1 | 0.1240 | 0.1255 | 0.1225 | 0.1237 | 0.1236 | 0.1238 |
| Module2 | 1.0    | 0.1061 | 0.1053 | 0.1070 | 0.1052 | 0.2847 |
| Module3 | 1.0    | 1.0    | 0.1091 | 0.1109 | 0.1104 | 0.4660 |
| Module4 | 1.0    | 1.0    | 1.0    | 0.1061 | 0.1038 | 0.6419 |
| Module5 | 1.0    | 1.0    | 1.0    | 1.0    | 0.1661 | 0.8332 |
| (a)     | 0.3114 | 0.4280 | 0.6677 | 0.7315 | 0.3760 | 0.5033 |
| (b)     | 1.0    | 1.0    | 1.0    | 0.9882 | 0.7061 | 0.9388 |
| (c)     | 0.9997 | 0.9996 | 0.9997 | 0.9886 | 0.1975 | 0.8370 |
| (d)     | 1.0    | 1.0    | 1.0    | 1.0    | 0.1115 | 0.8222 |
| (e)     | 1.0    | 1.0    | 1.0    | 1.0    | 0.4202 | 0.8840 |

(a). Majority voter using distance metric.
(b). Majority voter using bit-by-bit voting scheme.
(c). Adaptive majority voter.
(d). Adaptive voter with incoherence scoring.
(e). Dynamic adaptive voter with incoherence scoring.

*Some Remarks:*

State depended digital system such as controller circuitry, processors ..etc is quite vulnerable to errors occurrence in the combinatorial logic. In such case, error will sweep current states of system into a wrong state and cause completely failing of the system. For this kind of vulnerable system, error masking mechanism producing correct outputs should be deployed to obtain availability of 1.0 as much as possible. Adaptive voters are capable of producing correct output as long as there exist a fault-free redundant modules. Hence, adaptive voting can provide 1.0 availability longer than majority voters. This is why; adaptive voters are more convenient to be applied to state depended digital system than pure majority voting. Dynamic adaptive voting with incoherence scoring can exhibit availability of 1.0 until all of the redundant modules become faulty as seen in the Table 1. Other voter except adaptive voter with incoherence scoring lost 1.0 availability earlier. When all redundant modules became faulty, dynamic adaptive voting with incoherence scoring could detect the case of $(F,F,F,F,F)$ and it showed higher availability than none-dynamic version of it by relying on bit-by-bit majority voting results. When we look at total availabilities in the Table 1, it is seen that majority voter using bit-by-bit voting scheme has higher total availability. Unfortunately, it has availability lower than 1.0 in the case that there is still a fault-free module.

## c) Noisy Channel Test

Performances of voters were tested in noisy channel application. In this test, voters were fed from 5 noisy channels carrying the same sinus signal. Bit error rates (BER) at the output of voters listening these noisy channels were drawn versus number of the error bit at samples in the Figure 11. Amplitude of the sinus signal was represented by 8 bit samples. Error bits were applied on random selected bits of these samples. Test was done for 10.000 samples with error bits inserted up to 5 bits.

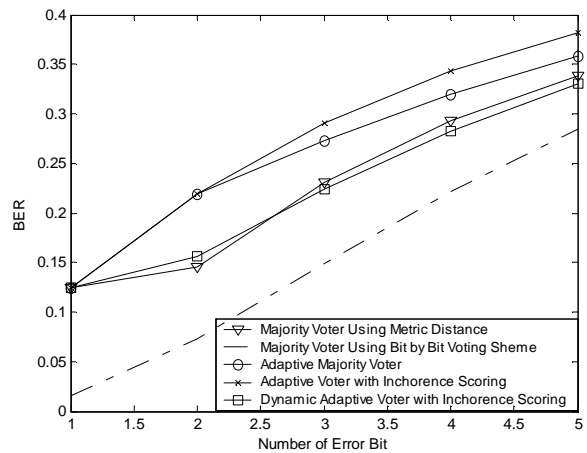

**Figure 11**. Voter's BER performance for the case that all channel are noisy

Expectedly, majority voter using bit-by-bit scheme has the lowest BER when all channels are noisy. Dynamic adaptive voter with incoherence scoring slightly better BER performance than majority voter using distance metric as a result of detecting that all redundant modules were faulty. Other voters mostly rely on history records of redundant modules and their BER is higher than majority votes.

## Conclusions

Availability of the voting methods is very conditional. In such case that, there is at least one fault-free module, adaptive voting can show higher availability. In the case all modules have similar level of error, majority voting is able to show higher availability than adaptive voting.

Adaptive voter with incoherence scoring is very aware of fault-free modules when its regulation parameters set to a low value. This property makes the voter reliable for the state depended digital system application, where perpetuating availability of 1.0 is crucial. Besides, its regulation parameter enables changing attitude of voter any time. Utilizing this property, dynamic version of adaptive voter with incoherence scoring could be developed. It was able to show better availability performance than static version of it.